# Applications of molecular communications to medicine: a survey

L. Felicetti, M. Femminella, G. Reali*, P. Liò

**Abstract**

In recent years, progresses in nanotechnology have established the foundations for implementing nanomachines capable of carrying out simple but significant tasks. Under this stimulus, researchers have been proposing various solutions for realizing nanoscale communications, considering both electromagnetic and biological communications. Their aim is to extend the capabilities of nanodevices, so as to enable the execution of more complex tasks by means of mutual coordination, achievable through communications. However, although most of these proposals show how devices can communicate at the nanoscales, they leave in the background specific applications of these new technologies. Thus, this paper shows an overview of the actual and potential applications that can rely on a specific class of such communications techniques, commonly referred to as molecular communications. In particular, we focus on health-related applications. This decision is due to the rapidly increasing interests of research communities and companies to minimally invasive, biocompatible, and targeted health-care solutions. Molecular communication techniques have actually the potentials of becoming the main technology for implementing advanced medical solution. Hence, in this paper we provide a taxonomy of potential applications, illustrate them in some details, along with the existing open challenges for them to be actually deployed, and draw future perspectives.

**Keywords:** molecular communications, nanomedicine, targeted scope, interfacing, control methods

L. Felicetti, M. Femminella, and G. Reali are with the Department of Engineering, University of Perugia, Via G. Duranti 93, 06125 Perugia, Italy. Emails: ing.luca.felicetti@gmail.com, mauro.femminella@unipg.it, gianluca.reali@unipg.it. Tel. +39 075 585 3651, fax +39 075 585 3654. P. Liò is with the Computer Laboratory, University of Cambridge, 15 JJ Thomson Avenue, Cambridge CB3 0FD, UK. Email: pl219@cam.ac.uk. Asterisk indicates the corresponding author.

# 1. Introduction

In recent years, nanoscale communications have inspired a huge research effort in many fields [1][2][3], including medical science [4][5], environmental control, and material science. Research activities have focused on different types of nanomachines, from those based on carbon nanotubes and carbon nanowires [6] to biological ones [7][8]. Due to the heterogeneity of environments and communication techniques exploitable at the nanoscales, it is unfeasible to identify general models, valid for most of nano-communication technologies [9], such as terahertz communications [10], neuronal communications [11], and molecular communications [12]. In fact, context and micro-environmental features have a deep influence on the models of each communication scenario. For this reason, an extensive literature relevant to the analysis of different innovative communication solutions at the nanoscales exists.

Molecular communications have received a lot of attention since they are considered an alternative approach to electromagnetic communications due to their unique features of biocompatibility and minimal invasiveness, which are essential for them to be used in living bodies. This class of communications takes inspiration from some existing communication mechanisms between biological entities. It consists of using relatively small molecules, such as hormones and other small proteins (e.g. cytokines), peptides, carbohydrates, lipids and combinations of them, which can propagate from a transmitter to receivers. The response of such communicating biological entities (i.e. natural and/or artificial cells) is highly specialized. Indeed, receivers must recognize different signals associated with different molecules received through their specialized membrane receptors. Actually, the cell membrane may contain hundreds or thousands of receptor molecules for each type of compatible ligand. Depending on the signals (i.e. types of molecules) received by a cell, different behaviors are triggered [14]. By exploiting these communication mechanisms, the ongoing research on molecular communications is highly oriented to expanding capabilities of nanodevices, by enabling the execution of complex tasks through their coordination. However, many proposals are essentially focused on design and analysis of communications techniques and protocols at the nanoscales, often without a sound analysis of the wide set of applications that can benefit from these proposals. The essential contribution of this works is to identify and classify the applications that can benefit from nanoscale molecular communications in the medical field. A contribution on this subject can be found in [4], although at a very early stage. It is also worth to mention the paper [13], which is a survey of nanotechnology applications for health care, but it does not include any discussion on the potential applications and benefits achievable by the introduction of molecular communications. Hence, this survey contributes to organize and classify the medical scenarios that are believed to significantly benefit from this kind of communications. Although full-fledged artificial biological nanomachines are far from being considered ready for systematic production and exploitation [15], biochemists have achieved significant milestones, such as the creation of cell components, such as artificial ribosomes, which can be used for the artificial synthesis of proteins [7][8]. In addition, the theoretical and computational models of biological circuits enabling molecular communication have already been assessed [16].

Some realistic results achievable in the mid-term consist of emulating nanoscale biological processes through computational intensive simulation platforms [17]. This way, it is possible to realize personalized predictions of the evolution/trajectories of diseases, starting from a limited number of biomarkers[1], without the need of executing the traditional in-vivo tests on patients [18][19]. The computational models are based on a detailed characterization of the molecular communication parameters done by using results of in-vitro and in-vivo experiments. The key aspect of this approach is translating general experimental results in numerical functions that can be parameterized through marker values relevant to each individual patient, so as to produce a personalized medical characterization of a patient, including risk assessments, thus helping medical personnel to identify optimal treatments.

An example of longer-term objectives of nanoscale communications consist of designing and implementing nano-sensors and nano-actuators for the prevention, detection, and treatment of a large set of diseases, such as cardiovascular diseases or tumors. For instance, specific proteins can be delivered for activating the immune system and/or triggering drug delivery systems in small specific areas without affecting the rest of the body. In addition, real-time monitoring of biological parameters during the execution of these innovative techniques can be integrated by an information exchange with the outer world. In this respect, a fundamental research activity consists of the design of communication protocols handling both intra-body communications and information exchange with the external through special interfaces. Recent advancements on this subject has brought insights on the development of body area nanonetworks able to make localized or remote decisions on the treatment of complex diseases [4].

Figure 1 shows a taxonomy of the main applications in nanomedicine supported by ICT. The first distinction is between techniques oriented to diagnosis and those oriented to treatment. Application developed for diagnostic purposes are further classified in disease detection applications, personalized applications for a detailed diagnosis after the initial disease detection, and advanced imaging techniques. For what concerns treatment applications, we have identified applications designed for drug delivery, tissue engineering, nanosurgery, and controlled triggering of the portions of the immune system. In addition, some applications can be further specialized by functions, such as run time monitoring, control, and activation/delivery, and be embedded in a temporal logic. Clearly, boundaries between diagnosis and treatment are blurring (predictive and preventable medicine), and the two fields overlap. However, this taxonomy allows associating the suitable technology with each application field.

The paper is organized as follows. Section 2 focuses on diagnostic applications. Different monitoring techniques for both general biological conditions and specific diseases are illustrated. In Section 3, we describe applications of nanocommunication systems for advanced treatments. In Section 4, we present some implementation mechanisms and interfaces currently under study. Section 5 draws future research challenges of molecular communications for medical purposes, followed by our conclusions in Section 6.

---

[1] A biomarker is a traceable substance that is introduced into an organism as a means to analyze some health-related aspects, or a substance whose detection indicates a particular disease state. In this work, we will refer to biomarkers as the second definition. Specifically, a biomarker indicates a change in expression or state of a protein that correlates with the risk or progression of a disease, or with the susceptibility of the disease to a given treatment.

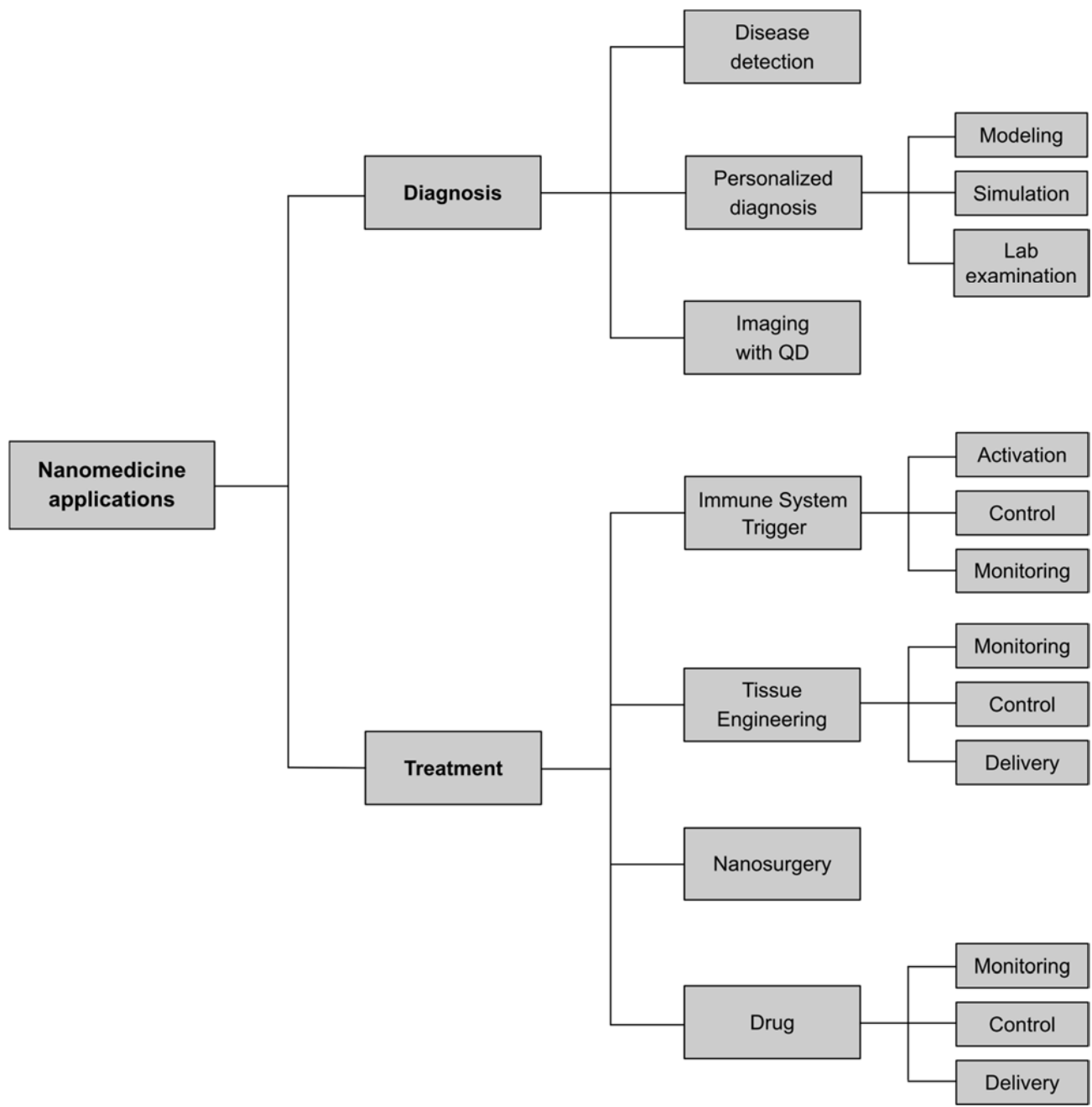

Figure 1: A taxonomy of potential applications of molecular communications to nanomedicine.

## 2. Diagnostic Applications

In this section we illustrate some applications proposed for diagnostic purposes. We focus on disease detection in subsection 2.1, in particular on tumor detection. Then, in subsection 2.2, we explore the field of personalized diagnosis, where the combination of advanced simulation techniques and limited clinical exams allows implementing personalized health monitoring tools and personalized determination of risk assessment. Finally, in subsection 2.3, we focus on the imaging techniques and show how their accuracy can benefit of the deployment of molecular communications techniques.

**2.1. Disease Detection**

A promising application area for the nanoscale communications is that of the diagnosis of diseases. For instance, the identification of tumors is particularly well suited to be achieved through molecular communications [20][21]. The basic idea consists of monitoring the concentration of specific biomarkers in the cardiocirculatory system. This way, it is possible to detect different cancer types in their early stages. Clearly, this approach is not limited to tumors and can be successfully applied to other critical diseases.

Some proposals consist of the development of nanorobots, which can be dragged towards the tumor site by flagellated bacteria attached to them, since the latter can be attracted by some proteins secreted by tumors. A graphical representation of this application is shown in Figure 2. A further step consists of making such nanorobots able to release a given amount of medicine directly on the desired location (drug delivery), or release signals for other bionanomachines, which have to be taken out of the body (see Section 4). The chemotactic capability of these bacteria induces a higher migration speed toward tumor cell lysates than toward normal cells. These results have been verified through a set of in-vitro and in-vivo experiments [22]. It is also possible to take advantage of this ability for monitoring the spread of the tumor cells and to implement a targeted drug directly on diseased cells thus reducing side effects [22]. For this purpose, it was also proposed to attach a set of polystyrene microbeads on the bacterial surface [23][24][25]. In this way, it is possible to develop various types of bacteria-based microrobots by using compatible materials able to bind with the polystyrene beads.

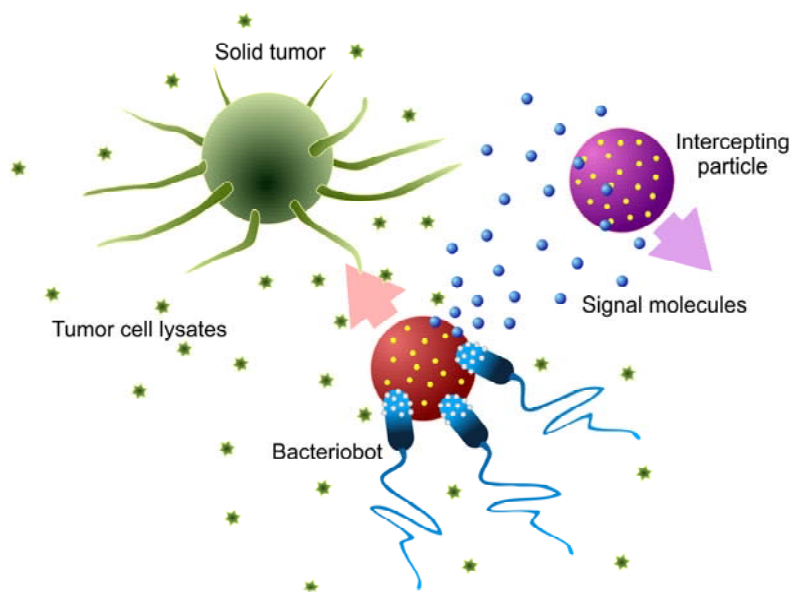

Figure 2: Nanorobots dragged towards the tumor site by flagellated bacteria attached to them; bacteria are attracted by some proteins secreted by tumors.

Another major strand of research is the detection of tumors by sensing abnormal concentration of circulating tumor cells (CTC) in the blood stream, that contribute to the spread of tumors throughout the body. Whilst in healthy immune conditions a minimal presence of CTC may be found in the blood stream, and the immune system is typically able to remove them, when a cancer has started to develop, CTCs originate massively from the primary tumor site and, through the bloodstream, may propagate in any part of the body, generating the so-called metastases if suitable conditions are found. Hence the concentration of such cells in the bloodstream gives useful diagnostic and prognostic information about the location and progression of a tumor. The monitoring of the CTC concentration allows both an early detection of a disease both in its initial phase and any at any relapse of it after an apparently successful treatment. Each CTC exposes several biomarkers on its surface that are useful for the cell detection. One of the most important biomarker is the CD47 that is in general over-expressed by the tumoral cells to fool the immune system and avoid to be destroyed by macrophages [26][27].

Other biomarkers can be used for detecting CTCs include the circulating microRNA-101 for the hepatocellular carcinoma [28], the carcinoembryonic antigen (CEA) for different types of lung cancer [29], the CD164 protein for the ovarian cancer [30], the apolipoprotein C-II for the cervical cancer [31], the plasma osteopontin for non-small cell lung cancer [32], and the plectin for the pancreatic cancer [33]. All these markers are expressed over the surface of CTCs, which are slightly larger than white blood cells (they seem also to detach from the primary tumours in groups). It may also happen that CTCs spread their DNA throughout the circulatory system by using microvesicles/exosomes as transport vector [34]. By the simultaneous detection of different types of biomarkers it is possible to identify different cancer types and their stage with a high degree of accuracy, thus obtaining also information about the location of the tumor. It is likely that risk ascertainment methodologies will rely on networks of markers or markers that will undergo remodulation in different progression of the disease.

Molecular communications have a central role for real-time detecting CTCs by using implanted devices. Detection can happen by either using contact-based communications [35], or by absorbing microvesicles/exosomes transporting RNA-i strands of the tumor emitted by CTCs [36]. Since the concentration of these cells could be very small in the early stages of a disease, it is necessary to collect some evidences in order to reliably estimate a higher than normal concentration. For increasing the detection probability, some proposals are based on the use of circulating mobile nano-sensors. For each individual CTC detection, they are designed for releasing bursts of signaling molecules that are collected by a sink, likely located in an implanted device in the circulatory system. In turn, this sink (e.g. the smart probe presented in Section 4) can deliver the collected information to the external by means different communication techniques. A high level illustration of this approach is shown in Figure 3. An alternative, although more complex solution, consists of processing the collected information locally, in sink nano-controllers or even in the detecting nano-sensors. We note that further improvements could rely on detection of both CTCs and free tumoral DNA in the blood.

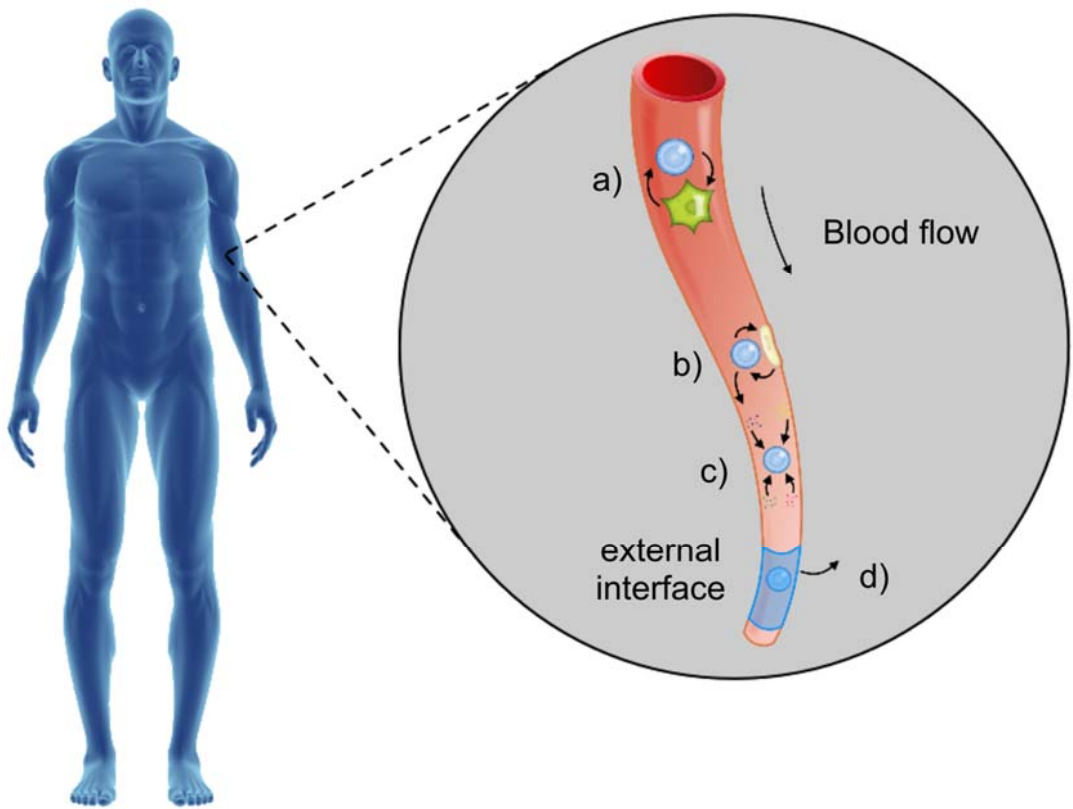

Figure 3: A schematic illustration of CTC detection assisted by molecular communications: a) nano-sensor-CTC interaction; b) nano-sensor interacting with a damaged vessel and/or fixed cell (e.g., a tumor); c) nano-sensor/nano-controller collecting bursts of signaling molecules released by cells or other nano-sensors; d) nano-sensor interacting with a smart probe to export collected information.

A further approach for (liver) tumor diagnosis, which makes use of molecular communications outside of the cardiocirculatory system, is to engineer microbes that are well tolerated by patients and that can be integrated with existing clinical methods. In [37], the authors make use of the safe probiotic Escherichia coli Nissle 1917 to develop an orally administered diagnostic that can indicate the presence of liver metastasis by producing easily detectable signals in urine. In fact, the used bacteria generate a high-contrast urine signal through selective expansion in liver metastases and trigger the production of a small molecule that can be detected in urine. The used bacterium colonizes tumor tissues in rodent models of liver metastasis without colonizing healthy organs or fibrotic liver tissue. The results in [37] demonstrate that probiotics can be programmed to safely and selectively deliver synthetic gene circuits to diseased tissue microenvironments.

Another approach which takes advantage of molecular communications for tumor detection consists of using a smart sensor chip for improving efficiency and accuracy of prostate cancer detection [38]. This approach do not involve the cardiocirculatory system, but leverages molecular communications in external devices. The sensor chip illustrated in [38] can both pick up differences in glycoprotein molecules and help early stage cancer diagnosis. Glycoprotein molecules play a key role in the immune response, thus they are useful biomarkers for detecting prostate cancer and other diseases. The sensor chip in [38] includes synthetic molecular communications receptors of specific glycoprotein molecules that are differentiated by their modified carbohydrate chains.

**2.2. Personalized Diagnosis**

Through the combination of interdisciplinary expertises, coming from the medical, engineering, and computer science research fields, it is envisaged the development of an innovative prevention approach based on a personalized risk assessment procedure. Indeed, due to the availability of technical information through the internet, the advancement in the medicine field has led also to an involvement of the patients in the health care process in order to give them a unique knowledge of their own health status[2]. This knowledge will let each patient to take part in each step of the therapeutic process and allow the delivery of information between medical staff, promoting the medical coordination and reducing possible errors. On this context, the model of P6 medicine [19], initially proposed by Leroy Hood as P4 (Predictive, personalized, preventive, participatory) model [18], has achieved a substantial consensus. P6 medicine is defined as *participatory*, *personalized*, *predictive*, *preventive*, *psycho-cognitive* and *public*. The strategic objective of a personalized diagnosis assisted by molecular communication is the definition, design, implementation, and experimental assessment of a not invasive and radically new method for assessing the risk of several diseases in the human body (e.g. cardiovascular diseases and many others).

In this context, solutions for monitoring of the evolution of the atherosclerosis, by dealing with the low-level mechanisms that govern its formation (atherogenesis) and its evolution, can be found in [39] and [40]. They can clearly be extended to other pathologies, such as diabetes. Essentially, extensive usage of massive computational resources and theoretical models are used for emulating the behaviour of cells (mainly platelets and white blood cells) and nanoparticles (mainly CD40L [41]) circulating in microcirculation, interacting both with each other and with endothelial cells of the blood vessels and responsible to be the cause of atherogenesis [42][43]). Through the design of specific computational methods it is possible to simulate the mechanisms leading to the development of atheroma within the inside lining of arteries and their subsequent hardening. A further development could be the in-silico emulation of the process of artery narrowing, which can have a major effect on blood flow, ultimately leading to the total closure of vessels that may cause diseases such as, angina, myocardial infarction, stroke, and critical limb ischemia, depending on the site of the atheroma. Clearly, the in-silico emulation has to be previously tuned by means of the analysis of the outcomes of a number of in-vitro/in-vivo experiments [44][45][40], in order to validate the simulation platform, and to make it so robust to be easily parameterized by means of a limited number of parameters, which can be extracted by standard lab examinations. Figure 4 shows the mapping of traditional communications systems elements onto the biological ones [45].

---

[2] Here we note that data ownership will be a debatable topic.

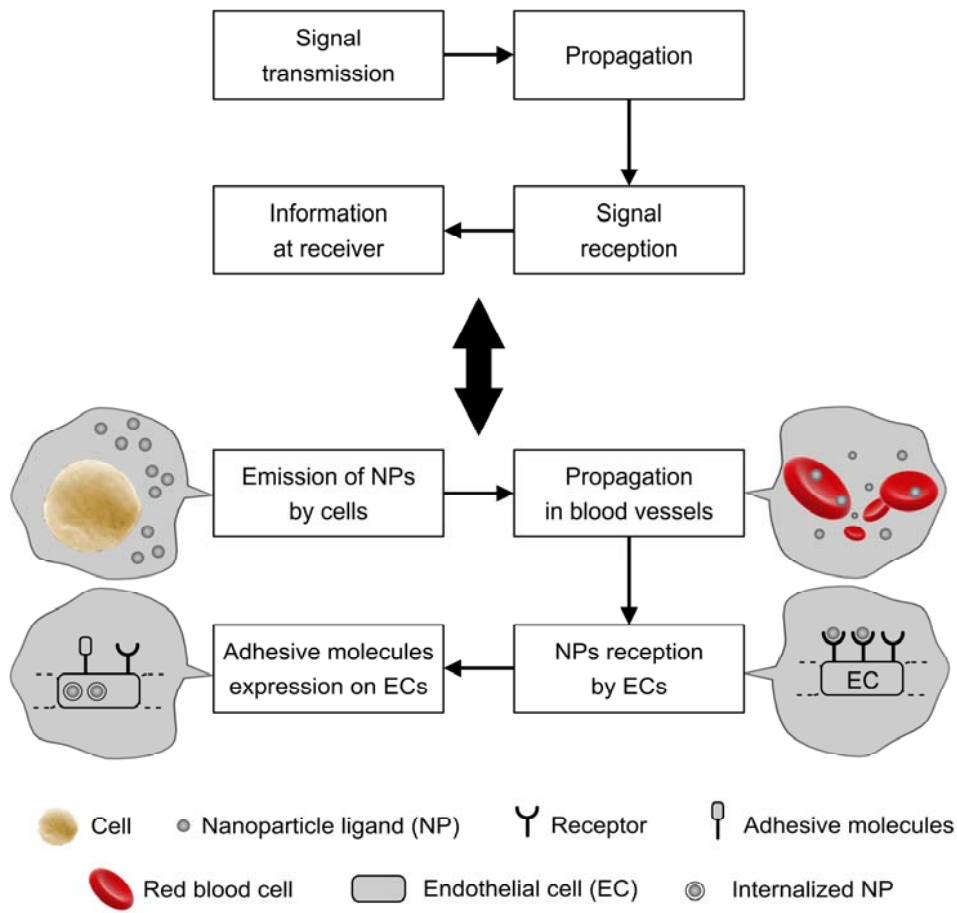

Figure 4: Mapping of the molecular signaling system in blood vessels onto the telecommunications counterpart.

The potentials of this approach consists of higher capabilities of determining one individual's actual cardiovascular risk if compared to current approaches with the final aim to increase the awareness of the actual risk of each individual subject so as to deploy suitable and personalized medical treatment that will bring to significant benefits in terms of:

- reduction of mortality and morbidity due to cardiovascular diseases;
- possibility of reducing time-consuming and potentially invasive cardiovascular diagnostic investigations and thus to improve the quality of life of patients;
- reduction of iatrogenic disease due to unnecessary/inappropriate preventive therapies;
- significant decrease of national health service expenditures.

From the medical perspective, it is necessary to carefully identify and measure all the significant biological and pathological processes involved in the development of atherothrombosis, and in particular the complex cell-cell interactions that take place in vivo during atherogenesis and arterial thrombosis, and to assess, their mutual influence and evolution [39]. Clearly, understanding which molecular communications take place, at which extent, and with which impact on atheroma formation, would be a major issue in this process.

The approach proposed in [39] aims to translate these processes in algorithms and to organize them within a suitable comprehensive simulator. The complexity of this simulator, although necessarily high, must allow its management through innovative ICT theories and technologies, and will certainly involve the use of the most powerful currently available solutions of distributed and cloud computing.

### 2.3. Imaging

An interesting nanoscale communication application is the visualization of biological parameters, based on fluorescent molecules, i.e. fluorophores, as transmitting and receiving nanomachines. These special molecules can be excited by optical, electrical, chemical or biological energy, and individually relax to the ground state after a random time. The excited-state of the donor can be suppressed in many ways, switching off the fluorescence. For example, a collision with excited donor may cause the quenching of the fluorescence, dissipating the energy through heat without the emission of photons.

In particular, quantum dots (QDs), which are tiny light-emitting particles at the nanometer scale, are emerging as a new class of fluorescent probes for in vivo biomolecular and cellular imaging. In comparison with organic dyes and fluorescent proteins, QDs have some advantages due to optical and electronic features: size-tunable light emission, improved signal brightness, resistance against photobleaching, and simultaneous excitation of multiple fluorescence colors. QDs have opened new possibilities for ultrasensitive and multiplexed imaging of molecular targets in living cells, animal models and, possibly, in humans [46], with increased accuracy compared to the classic imaging techniques.

A deeply analyzed quenching mechanism is the Förster resonance energy transfer (FRET), which is based on the migration of the excited-state energy, in the form of exciton, from the donor molecule to a ground-state acceptor molecule. In this case, the donor does not fluoresce, and the acceptor non-radiatively becomes excited by receiving the donor's exciton [47]. The FRET approach has inspired a sort of short-range communications, in which the information are encoded into the excited-states of the fluorophores, and then are transferred via FRET mechanism between a donor fluorophore and an acceptor fluorophore, one acting as a transmitter antenna and the other as a receiver one.

Through the excitation of the fluorophore by a laser, with a wavelength in the range of donor's absorption spectrum and using the on-off keying modulation scheme, it was analyzed also the channel capacity [48]. The simulation results showed that it is possible to achieve transmission rates up to 10Mbps over short communication ranges, [49]. In [50], the authors propose a FRET-based mobile molecular sensor/actuator network (FRET-MSAN) composed of a set of sensors, i.e. bioluminescent molecules that are excited upon binding a target molecule and actuators, which can realize a specific task after proper excitation.

In general, bioluminescent molecules do not require an external excitation source, since they may react with endogenous elements, such as calcium ions, and release a photon. For example, in [51] Quantum Dots-

DNA (QDs-DNA) nanosensors based on FRET are used to detect target DNA and single mismatch in Hepatitis B Virus (HBV) gene.

An interesting active targeting solution is given by the aptamers, known as "chemical antibodies" due to their functional similarities [52]. The aptamers are composed by single-stranded RNA or DNA nucleic acids that allow the recognition and binding to specific targets with a high degree of specificity and affinity. Moreover, they have higher thermal stability, faster tissue penetration, low-cost chemical synthesis than their natural counterpart. Targeted imaging is a possible application of aptamers. In fact, they could be combined with imaging nanomaterials, radionuclide probes (e.g. magnetic nanomaterials used in magnetic resonance imaging) or even QDs to reach a specific target and improve accuracy of imaging. Clearly the molecular imaging (for example intravital) together with the medical imaging will provide impressive capability to the diagnosis.

Another interesting proposal for smart imaging consists of the so-called "virtual biopsy" [53]. It is implemented through a biodegradable, nano-sized polymer, which is armed with tracers that detect cancer biomarkers. Visible on a magnetic resonance imaging test, tracers accumulate at the cancer site and diagnose the type of cells (e.g. breast). After this "virtual biopsy", a drug delivery system, equipped with drugs tailored to the specific mutations, can deliver them to the tumor (see section 3.1). The authors claim that this technique can be applied also to brain tumors [53], which are notoriously difficult to treat. In particular, the molecular structures of most chemotherapy drugs are too large to penetrate the blood-brain barrier, which is the system of protective blood vessels that surround brain.

## 3. Treatment of Diseases

One of the most interesting class of applications of molecular communications is the detection and treatment of diseases, with special focus on tumors. Further applications on health care include immune system triggering, with applications not only to tumors, but also to other localized diseases (e.g. inflammations), tissue engineering in regenerative medicine, and nanosurgery.

### 3.1. Drug Delivery

One of the most popular topics in nanomedicine is the targeted drug delivery, since it could be the basis for the modern medical therapeutics. Its main goal is to provide a localized drug delivery only where medication is needed, thus avoiding to affect other healthy parts of the body. This is due to the very small size of the particles delivering medicine, which is in the order of nanometers, which allow them to diffuse into the bloodstream and across the vascular and interstitial barriers.

Many types of drug delivery, based on the molecular communications, have been proposed in the technical literature, with both passive and active transport of molecules. For what concerns passive transport, the most popular approach is based on particle diffusion, with or without drift. In case the drug delivery happens via the cardiocirculatory or lymphatic system, the most appropriate model is the diffusion with drift, by also considering the effect of collisions with blood cells [40][39]. In case the drug delivery happens outside blood vessels, e.g. in the extracellular matrix, the mere diffusion-based model is sufficiently accurate. In case of active transport, the main alternatives are those based on bacteria or molecular motors.

In order to design an efficient and targeted delivery, it is essential to correctly determine the distribution of particles over time. For this reason, the relevant research has produced several approaches, from statistical models [54][55] to analytical ones [56]. In particular, drugs delivered through bloodstream can reach any part of the body, by passing through the complex network of blood vessels (Particulate Drug Delivery System, PDDS). In [56] the authors propose a specific model of the cardiovascular system, by analyzing the peculiarities of different vessels, from the smallest to the largest ones, thus obtaining a drug propagation network model, which takes into account even bifurcations and junctions of vessels. The resulting model includes also the individual features of the cardiovascular system by physiological parameters, such as the heartbeat rate profile. This model could be useful to guarantee that the drug concentration does not reach toxic levels in the body and it is a essential for optimizing the drug injection, by identifying the suitable injection points that guarantee the best delivery profile for maximizing the treatment benefits while minimizing the amount of drugs in the other parts of the body. To this aim, it is necessary to realize a complete control of the drug release, from the initial release, to the suitable release rate.

A theoretical model on the local rate control between a nano-controller and a nano-actuator, not involving drug propagation in the cardiocirculatory system (local drug delivery system), has been presented in [57]. Subsequently, a related communication protocol, inspired by the TCP congestion control, has been proposed in [58]. This protocol can control the drug release process by using feedback messages. A challenging aspect of the protocol operation is that feedback messages, in some cases, can be transmitted even in situation where a drift of the communication medium is present. A detailed discussion of the conditions that enable this type of communication is presented in [59].

Even real time monitoring of the drug concentration is very important. Indeed, as illustrated for CTC monitoring, a set of nanomachines able to recognize specific drug molecules could take decisions based on the measured levels of their concentration. This way, it is possible to control the drug release rate and, at the same time, to deliver the collected information outside, through specific interfaces, as illustrated in Section 4.

In synthesis, we can identify three main challenging functions, and associated entities, related to drug delivery, (see Figure 1):

- *Delivery*: It is the function of releasing drugs, realized by nano-actuators. Delivery can be either systemic, as proposed in [56], or localized, as mentioned in [57][58]. In the first case, in order to avoid side effects, a preliminary configuration and provisioning of the system is needed to regulate the amount of drug to release. For realizing a localized delivery, additional entities are necessary, as explained below;

- *Monitoring:* It is the function monitoring the release rate of the drug for it to be effective. It can be implemented by a monitoring device communicating with the target (e.g. a tumor), or by deploying on a monitoring device, located near to the target, some receptors able to absorb part of the drug. Depending on either the absorption rate measured by such a monitoring node, or evaluated by processing the information received through a molecular communication channel with the target (e.g. via contact-based communications), the control node can take appropriate actions. In fact, an excessive amount of drug should be avoided not only for its undesirable side effects, but also for its possible wastage, as it happens in case of receiver/channel congestion [57][58];
- *Control:* the control node is a nano-node in charge of regulating the drug release rate at the nano-actuator. It can be implemented as a separate nano-node, or co-located with the monitoring node. It can trigger variations in the release rate of the nano-actuator by means of feedback messages, transmitted by using molecules different from those of the monitored drug [57][58]. The control node is also in charge to start and stop the drug release, upon specific local conditions, thus implementing a sort of connection-oriented communications [58].

The functions mentioned above and their implementation by means of molecular communications are illustrated in Figure 5.

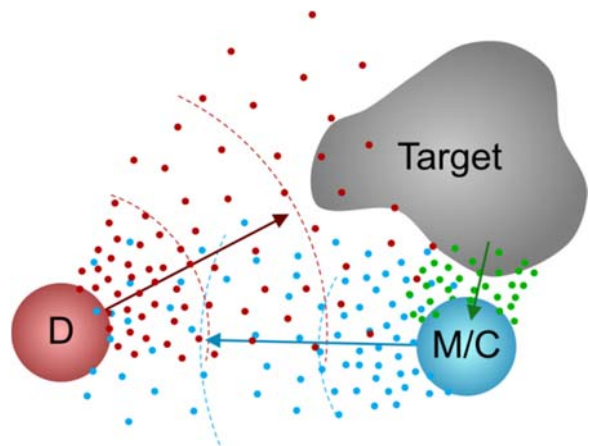

Figure 5: A general model of a drug delivery system: *D* is the actuator nano-node, in charge of delivering drugs (red particles); *M/C* is a monitor and/ control nano-node, in charge of measuring the effects of drugs on the *Target* by measuring specific quantities (green particles), and to control the emission rate of *D* via feedback signals (blue particles).

### 3.1.1. Virus-based techniques

A novel strand of research encompasses the use of virus particles as information carriers for molecular communication [60]. These organic particles consist of a nucleic payload (DNA or RNA), a protein coat that protects the information payload, and a lipid envelope that encapsulates the protein coat. This way, it is possible to exchange nucleic acid-based messages between nanomachines within biological environments. The protein coated which protects the payload may be regarded as the header of an information packet, since it implements a ligand compliant with the relevant receptors on the surface of the destination cells. This behavior is inherited from the basic features of the viruses, which makes them very useful in a nanoscale communication system. Indeed, each virus can encapsulate nucleic acid molecules and release them at long distances directly into the recipient cells. This property makes viruses useful for gene therapy applications. In order to increase the propagation distances, viruses can release the carried information into an intermediate cell, and in turn the infected cell spreads the infection acting as a relay node. The information encoded into the nucleic acid molecules contains the virus genetic information, such as the replication instructions for the infected cells.

In [60] the authors propose the use of synthetic cells as intermediate nodes that have been engineered so as to forward messages to the desired destination nodes, through multihop communications, thus implementing a sort of multihop routing technique. The source cell contains several payload genes, which allow transmitting multiple payloads destined to multiple destinations by using a single-source cell. Through the use of a gene expression control, it is possible to silence specific aspects of the viral gene expression, thus allowing the control of proviral genes that are responsible for payload selection, packing, and addressing of viruses in the cell [61].

By taking advantage of the abilities of viruses, it is possible to design complex networks where the released messages are received only by the desired nodes, in any desired directions. The viral vectors can therefore be used to deliver artificial payloads (i.e. DNA plasmid) into the targeted cells, thus realizing an effective treatment of several diseases [62]. This method allows packaging drugs and delivering them with great benefits in terms of preventing both drug degradation and interaction with the biological environment, and enhancing drug absorption on the desired tissues.

The paper [63] presents another technique to enhance drug absorption, even prolonging its interaction with the targeted tissue. Many drug types, that can be attached to a virus structure through a covalent binding, have been analyzed. Indeed, when viruses are used as nanocarriers, the targeting functions can be realized by chemically attaching the targeting domain either via bioconjugation to the exterior of the virus particles or by the genetically inserting the sequence of a domain into the surface loops of the viral capsid protein. Moreover, viruses have a regular structure and, for this reason, their chemical and conformational structure can be produced in large quantity with a high degree of accuracy.

### 3.1.2. Bacteria-based techniques

Molecular communications making use of bacteria can be used for executing all the three basic functions mentioned in Section 3.1, with a particular focus on cancer detection and treatment. In more details, the set of combined bacteria-based detection and treatment techniques include circulating tumor cell (CTC) detection in the blood stream, injection of genetically modified bacteria in tumor mass, real time monitoring of tumor evolution and bacteria transcriptome, proteome, and concentration in time and body organs, and counteraction of inflammations due to treatment.

It is known that tumors include different areas with very different concentration of oxygen, necrotic, hypoxic, and well-oxygenated regions. Due to the lack of oxygen, tumors with hypoxic areas are resistant to well-established anticancer agents and radiotherapy. On the other hand, the oxygen shortage can be exploited to allow anaerobic bacteria to grow inside a tumor. In particular, recent studies show the usage of an attenuated strain of the anaerobic bacterium Clostridium novyi. What emerged is a clear and reproducible antitumor response when the bacterium spores are injected into tumors. Experiments on rats, dogs, and even one human patient have been executed [64]. The achieved results are such to encourage a deep research on this direction. The idea of using bacteria to combat cancer dates back to the 1890's, when cancer researcher William Coley noticed that some patients who developed postsurgical infections went into remission or were even cured of their disease.

A recent research has demonstrated the feasibility of a cancer therapy based on genetically modified bacteria [65]. Experimental results show that bacteria are highly selective and localized within tumors [65], since an anaerobic microenvironment is ideal for their growth [66]. The study reveals that the modified bacteria prefer to replicate in tumors. Moreover, the modification allows the expression of a variety of genes coding for therapeutic proteins, such as tumor necrosis factor (TNF-alpha), and many others [67][68]. The molecular communications associated with the release of such proteins have shown a remarkable tumor regression with respect to the untreated tumors with such bacteria [65].

Although bacteria were genetically modified, some undesired side effects appeared and the effectiveness of this approach was questioned. A deeper research activity is needed in order to reduce the threat of bacteria and their specific goal, by targeting only the tumor tissues and not anything else.

A further stand of research aiming at developing bacteria-based nanomachines able to accomplish diagnostic test, delicate surgeries, and cancer therapies, has brought to the design of magnetic drug carriers that can be guided across the blood vessels [69]. The tiny size of these nanomachines makes it difficult to implement a battery-powered motor. Thus, the guided propagation is obtained by equipping nanorobots with metallic particles that allow driving them by an external magnetic field (e.g. by a MRI machine). Such nanomachines can be obtained by using the MC-1 bacterium, which is a perfect carrier for anticancer drugs, that are attached to its external surface. Another important feature of this family of bacteria, is that they are magnetotactic, since they have the ability to align themselves with Earth's magnetic field. This allows guiding bacteria by generating a weak magnetic field. The release of such bacteria could be done through bubble-like vesicles that carry large amounts of bacteria and magnetic nanoparticles, in order to guide them towards the site of interest, where bacteria are released.

An open issue related to the use of bacteria is the effectiveness of their communications, especially when transmitted messages have to drive their operations. A technique that can improve the information delivery in bacterial nanonetworks is conjugation. It consists of forming a physical connection between bacteria in order to transfer DNA molecules (i.e., plasmids or chromosomes). However, the fragile physical connection between bacteria is prone to breakage, in particular under mechanical stress. In [70], a coding process for improving information delivery in bacterial nanonetworks is proposed and analyzed.

### 3.1.3. Nanorobots

An interesting approach for developing nanoscale-computing machines involves the use of DNA origami nanorobots [71]. This system can retrieve biochemical and physiological information for diagnostic purposes, by monitoring the fluorescent outputs associated with the robot final states. Moreover, the system is particularly useful for controlling therapeutic molecules in living organisms.

The main characteristics of these robots is that they can execute these computing functions (e.g. monitoring) acting as a network of logic gates. In particular, they are controlled by a gate that opens in response to a specific combination of proteins. This way, robots undergo a conformational change, expose their inner payload and make it available to engage target cells. The gate can be opened also by an external DNA strand that eventually activates the robot. This DNA strand (or key) can be loaded into one robot as a payload, and, if it is activated, the payload generated by this robot can enter the gate of another robot, thus activating it. These keys can cause a positive regulation or a negative regulation on the target robot. In the former case, it may open the gate, otherwise it may close or inhibit the opening of the door [72]. The architectures proposed in [72] can process only two bits, but through a daisy chain of processors, it is possible increase the overall processing capacity. Thus, by properly combining nanorobots, it is possible to obtain the outputs as the basic logical gates (AND, OR, NOT, XOR, NAND) and a half adder [73].

### 3.1.4. Antibody-based techniques

Another promising therapeutic method is the antibody-based drug delivery system (ADDS, [74]), which takes inspiration from the human autoimmune mechanisms. Such mechanisms can perform self-diagnosis, and allow also destroying the root cause of a disease, in an adaptive fashion. In the approach proposed in [74], the antibody-based drugs are artificially engineered in order to act similarly as the natural antibodies [75]. By taking advantage of the capabilities of such molecules to bind only with the antigen exposed on the diseased cells, it is possible to avoid side effects on healthy cells.

In [76], the authors extend the PDDS [56] in order to analyze and model the molecular transport on ADDS. By modifying the PDDS framework, it is possible to take into account the shape of the antibody molecules in order to track the drug propagation with a high temporal and spatial resolution. This is useful for obtaining accurate pharmacokinetic and pharmacodynamics models of antibody-based therapies.

Beside the targeted drug delivery systems, it is also possible to implement stimuli triggered drug delivery systems by using aptamers, introduced in Section 2.3. A stimuli-triggered drug release system could both improve the effectiveness of therapies and reduce their side effects. When a specific stimulus is detected by these systems, drugs are released directly at the specific disease location. Some of such stimuli include specific pH and temperature values, ionic strength, redox reagents, specific disease targets, light, and magnetic field.

A pH-triggered aptamer drug delivery system is illustrated in [52]. This kind of stimuli-triggered feature could be exploited also by nano-control nodes, which could encode the control messages through the modification of the local physical conditions in the site of interest, instead of releasing molecular signals, for triggering drug release.

*3.1.5. QD and FRET-based techniques*

Special kinds of fluorescent molecules find application in the photodynamic therapy (PDT) of cancer as actuators. Specifically, in PDT based on quantum dots, an excited dot transfers its optical energy to a conjugated photosensitizing agent which, subsequently, synthesizes a reactive singlet oxygen via energy transfer [77]. Through a well-planned excitation of quantum dots, it is possible to produce oxygen near to cancer cells, thus triggering the apoptotic process that brings these cells to death. By such mechanism, it is possible to develop a network composed by molecular sensors and actuators, in which mobile bioluminescent sensors transfer the information to fluorophore-based actuators, which can perform the appropriate actions. In particular, nanosensors may generate luminescent messages in order to notify actuators of a threat (e.g. a tumor).

Recent advances have led to the development of multifunctional nanoparticle probes that are very bright and stable under complex in vivo conditions. A new structural design involves both the encapsulation of luminescent QDs with amphiphilic block copolymers and combination of polymer coating to tumor-targeting ligands and drug delivery capabilities. Polymer-encapsulated QDs are essentially nontoxic to cells and animals, but their long-term in vivo toxicity and degradation need more studies [46].

**3.2. Immune System Activation**

In recent years, the interactions between immune cells have inspired a strand of research on the use of such mechanisms for nanoscale communications. Indeed, the immune system is composed by several types of cells able to perform specific tasks in order to defeat pathogens and other threats. Each immune cell type is specialized against specific threats and can be triggered by a cascade of signaling molecules released by specific immune cells, specialized in the recognizing specific pathogens. The immune system triggers not only reactions aiming to eliminate bacteria and viruses, or to clean infected cells, but also specific actions to eradicate tumor cells or any foreign body that is incompatible with the defended organism. The fundamental mechanisms underlying the defensive phenomena are essentially two:

- the cell or cell-mediated immunity, which is formed by cells with phagocytic activity;
- the humoral immunity, made of antibody molecules produced by the immune system.

By means of molecular transmissions from artificial nanomachines, it was proposed to trigger the immune response in order to intensify the utilization of its fighting capabilities against specific threats [78].

In [17] the authors have analyzed a mechanism for activating the humoral response through the activation of the T-lymphocytes [79][80]. These cells release molecules that produce a positive feedback which amplifies the signal delivered to the B-lymphocytes, which release antibodies.

Conversely, in some situations it is necessary to inhibit the immune response, by releasing molecules that can reduce the immune activities, or mitigate the immune response, by hindering the carrier propagation and absorption by using artificial antagonists. Such antagonists, that could be released by nanomachines, affect the assimilation phenomena in three ways:

- The receptor antagonists can bind to them, thus making them busy with any further assimilation, without activating it. In this way, any excitation stimulus vanishes [85][86].
- Soluble receptor antagonists can bind to the activation molecules, making carrier binding with the compliant receptors impossible [82][83].
- A set of decoy receptors are used to capture the activation molecules, not allowing them to reach the target cells [84].

Similarly to drug delivery applications, also for immune system activation it is possible to identify three distinct groups of functions: actuator, monitoring, and control (see Figure 1).

## 3.3. Tissue Engineering

Recently, significant research efforts have been dedicated to some technologies that allow engineering of tissues, and can lead to organ construction [87][88]. If successful, given the actual shortage of organs, these technologies could provide alternative treatments for patients with organ and tissue failure. Nanomachines embedded inside such engineered organs may produce smart organs, which allow detecting diseases.

A very interesting and minimally invasive application in this area is the detection of biological functions in cells, tissues, and organs, by monitoring the calcium ions behavior. Indeed, short-range communications obtained through calcium signaling is one of the most studied diffusion-based communications, where information is typically encoded by modulating ion concentration [89]. In recent years, research has focused on communications between nanomachines that may interface with tissues within organs by means of calcium ions [90][91]. In [92], the authors analyze the effects that tissue deformations have on the calcium ions propagation. Indeed, this molecular exchange is very common in intracellular communications (e.g. see [93]) and the transmission reliability depends on the state of the physical tissue structure. A set of nanomachines embedded into a tissue may collect information about the current state of the tissue, and use the statistics collected from the communication between different nanomachines in the network. To this aims, the authors propose a molecular nanonetwork inference process in order to determine the quantity and type of deformation, the concentration of the released ions, and the communication distance between source and receiver nodes. Since the structure of the tissues is composed by a tight formation of interconnected cells, any change in the structure of a cell could influence the whole tissue. In fact, the propagation of calcium ions, which diffuse within these tissues, depend largely on the inner spatial volume of the cells that compose the whole tissue. Therefore, any physical perturbation that affects the cell volume can also affect the communication capacity.

Nanomachines envisaged for such a scenario could be either biological or artificial. For the last ones, artificially assembled nanodevices can be implemented by using the nano-electromechanical systems (NEMS) technology. These devices, embedded into a tissue, could externally activate the release of calcium ions by using needles or force probes [94][95]. Even receivers could be implemented by using NEMS devices.

The tissue engineering has a primary role also in the field of the regenerative medicine [96]. Its primary goals are the replacement of damaged tissues and organs by using recruited cells or transplanted cells directly on the damaged site. Cellular signaling is fundamental in the development of tissues. This signaling mainly consists of diffusion-based molecular communications, but also mechanical cues and direct cell-to-cell interactions are involved in such processes. In particular, the molecules involved (peptides and proteins), act as potential therapeutics, regulate the absorption of soluble growth factors, and the adhesion, the migration, the proliferation, and the differentiation of a multitude of cell types. All these elements are involved in tissue engineering [97], and allow controlling the response of cells so as to control the formation of new tissues. Even the transplantation of cells able to release the desired growth factors has been proposed [98].

The effectiveness of the growth factors on the tissue repair process need a further analysis. Indeed, it is known that such factors are crucial for the control of a multitude of biological processes, included the tissue regeneration. Nevertheless, they have a short life once released in the body, since they are rapidly eliminated [99]. Success of tissue regeneration depends on the exposure time to such factors. The highest regeneration rate is obtained by an exposure that lasts for the whole repair process. For this reason, a set of controlled delivery systems able to release the growth factors, have been developed in order to prolong the tissue exposure and to maintain the concentration of these factors nearly constant [100].

A specific scope for the tissue engineering is neovascularization, which is crucial for the treatment of many coronary artery diseases. Through a deeper study on the angiogenesis phenomena, it was possible to identify many growth factors involved. Anyway, angiogenesis depends of a multitude of signaling molecules (i.e. cytokines) that behave in a coordinated manner.

Even bone regeneration has been considered in order to accelerate the regeneration process and to control the such a process in situations of incomplete regeneration. Medical researchers have found that magnetic nanoparticles coated with targeting proteins (ligand for molecular communications) can stimulate stem cells to regenerate bone. The regeneration process is maintained through the release of growth stimulant [101].

The potential applications include also wound healing, which consists of regenerating the skin tissue after ulcers, scars or burns. The task coordination between the bionanomachines implementing these functions can be accomplished through molecular communications, which allow realizing the precise release and control in space and time of growth factors and cytokines that allow shaping of tissues and organs.

Similarly as was done for drug delivery applications and immune system activation, also for tissue engineering three distinct groups of functions/nodes can be identified: actuators (e.g. for delivery of grow factors), monitoring (e.g. calcium ions monitoring), and control (e.g. rate control of grow factor via release of proteins).

**3.4. Nanosurgery**

A set of coordinated nanomachines, equipped with specific actuators and instruments, was conceived to perform nanosurgery interventions from the inside of the patient's body [102]. These nanomachines can be managed from the external through compatible interfaces under the supervision of human surgeon, acting as a semi-autonomous on-site system correcting lesions by nanomanipulation.

In this way, high precision intracellular surgeries, which are unfeasible through direct human manipulation, can be accomplished. These devices could realize interventions such as removal of microvascular obstructions, reconditioning of vascular endothelial cells, noninvasive tissue and organ transplantation, reparation of damaged extracellular and intracellular structures, and even replacement of chromosomes inside human cells [103].

In addition, also atomic force microscopy (AFM) can be extremely useful in nanosurgery applications, by taking advantage of its bioconjugation capabilities coupled with active assembly and analysis of bioconjugation-based nanostructures [104]. In particular, in [103] the authors have proposed an AFM-based nanorobot with integrated imaging, manipulation, analysis, and tracking functions, important for cellular-level surgery on live samples. It is a sort of augmented reality system that includes the possibility of having a visual feedback for monitoring dynamic changes on a sample surface. This nanodevice was demonstrated to be able to both deliver epidermal growth factor (EGF) to a cell and subsequently measure the elasticity response of cells contacting the stimulated cell.

## 4. Implementation Mechanisms and Interfaces

An open research issue in molecular communications aims to design suitable interfaces able to interconnect the nanoscale biological environment with the external world (outside the body). Such interfaces are conceived for monitoring and controlling intra-body nanodevices by means conventional external devices.

One of the proposals [20] assumes the cardiocirculatory system to be the transmission medium. This proposal uses a number of mobile, circulating nanomachines, which act as sensors, and a smart, fixed probe, disposed along a blood vessel, which act as the sink for the information collected by the sensors. After the probe has received specific signals from the circulating nanomachines, it delivers their information content to the external, as depicted in Figure 3. Considering the drag effect of the blood flow and the relatively high propagation times required to complete a successful transmission, the smart probe should be able of capturing the circulating nanomachines and keep them stuck to its surface at least for the time required to complete the transmission. This is possible by exposing suitable adhesion molecules on the smart probe surface. In such a scenario, the smart probe can trigger the communication from the mobile nanomachines stuck on its surface through the release of activation messages. The smart probe may collect several messages coming from different circulating nanomachines before releasing the overall information to the outside. This information could be encoded by the intensity of the received messages. In particular, a large amount of activated nanomachines, emitting the same signal (e.g. detection of tumor cells), will rise the alarm level. Each detected phenomenon can be coded through a specific pattern of molecules, so as to discern between a significantly large of biological parameters and extract the desired information from their joint processing. The information stored on each mobile nanomachine is cleared after its release to the smart probe. On the other hand, the smart probe may decide to release the collected information to the outside after a given number of assimilations. The externally controlled smart probe could be also used to control nanomachines, by transmitting commands that modify their inner state and their behavior.

Another further proposal consists of the dermal display, which envisages the use of a large number of bionanomachines embedded just few microns below the epidermis. It is therefore possible to obtain an interactive display, a few centimeters wide, able to both respond to the touch of a finger and emit visible photons in order to display the desired information over the skin [105]. The finger touch is an external stimulus that triggers a mechanical signal able to activate the bionanomachines that, through the molecular communication, carry health information. More precisely, the system makes use of two kind of interfaces, the first one is used for nanomachines to nanomachines communication (required for nanomachine interaction and coordination), and the second one refers to the inbound and outbound messages through the skin. The inbound interface converts signals from the external into chemical ones, while the outbound interface converts the chemical signal generated by nanomachines (i.e. molecular signals) into conventional electrical or optical signals. As for the inbound signals, a set of special materials can be used for realizing the transduction process. For example, photosensitive materials can convert optical signals transmitted by external devices into chemical signals that can be received by nanomachines. For instance, specific wavelengths can activate photoactivable proteins [106], and caged-compounds release encapsulated molecules through bond breaking [107]. Even temperature-sensitive materials could be very useful, since they can convert external thermal signals into chemical ones. A temperature increase can induce a conformational change on temperature-sensitive liposomes and dendrimers [108], thus causing depletion of their inner structures, through the release of molecules. In [105] the authors propose an approach where artificial materials are embedded into the cell cytosol in order to use those materials as interfaces (ART: artificially synthesized materials). They are believed to give more predictable responses to incoming and outcoming messages, compared to a genetically engineered solutions. Figure 6 shows a scheme of the input-output architecture proposed in [105].

By means of specific interfaces, it was proposed to connect biological nanonetworks not only with external devices, but also to the Internet, thus realizing the so-called Internet of Biological Nano Things [109].

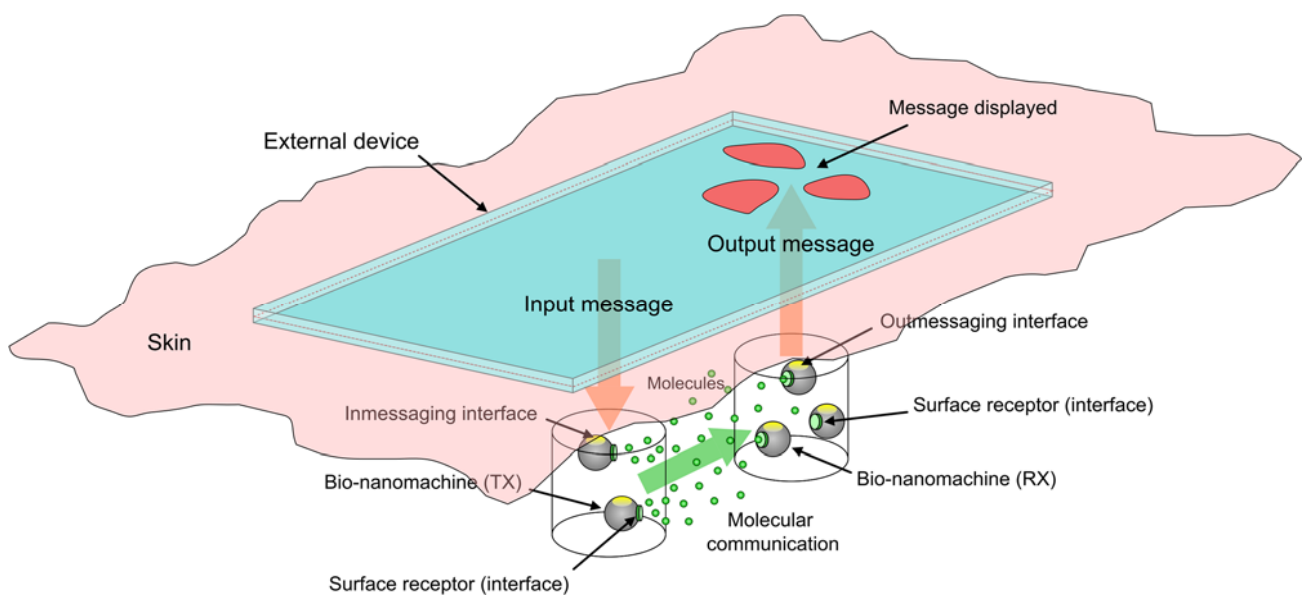

Figure 6: The architecture of the molecular communication system externally controlled by a dermal display

Finally, even magnetic materials are considered suitable signal transducers, since they can convert external magnetic signals into chemical ones. In this case, an external radio-frequency magnetic fields may cause reactions on gold nanocrystals attached to the DNA molecules and induce the hybridization/dehybridization process of the DNA molecules [110]. Bionanomachines integrated with these materials can react to the generated DNA molecules and thus respond to magnetic signals.

For the outbound interface, luminescent materials can be used for converting chemical (molecular) signals into optical signals to be sent to external devices. In this case, the bioluminescent molecules catalyze chemical reactions, through consumption of chemical energy (e.g. ATP), emitting light photons [111].

A recent study shown in [112] is focused on the use of sweat for diagnostic purposes. The main idea arises from the knowledge that sweat contains a huge amount of medical information, which is updated in real time. Indeed, since 1970s, several studies tried to use the sweat to monitor drug levels inside body. Taking inspiration from this, it was proposed a more controllable way to collect health information, by means the stimulation of perspiration using an engineered patch, which is also able to analyze the collected information. This patch is also capable of transmitting results to a remote device wirelessly (i.e. to a smartphone) thus creating a portable and easy to use monitoring system. It has been developed at the University of Cincinnati [112], and makes use of paper microfluidics to wick sweat from the skin through a specific membrane able to select only the desired ions. A specific circuitry is installed over the patch for both processing the collected samples and sending data to an external device. The onboard chip is externally powered similarly as in RFIDs. This kind of interface could be very useful for real time monitoring in different applications, such as those used in athletics and military fields. Figure 7 illustrates the layout of the engineered patch.

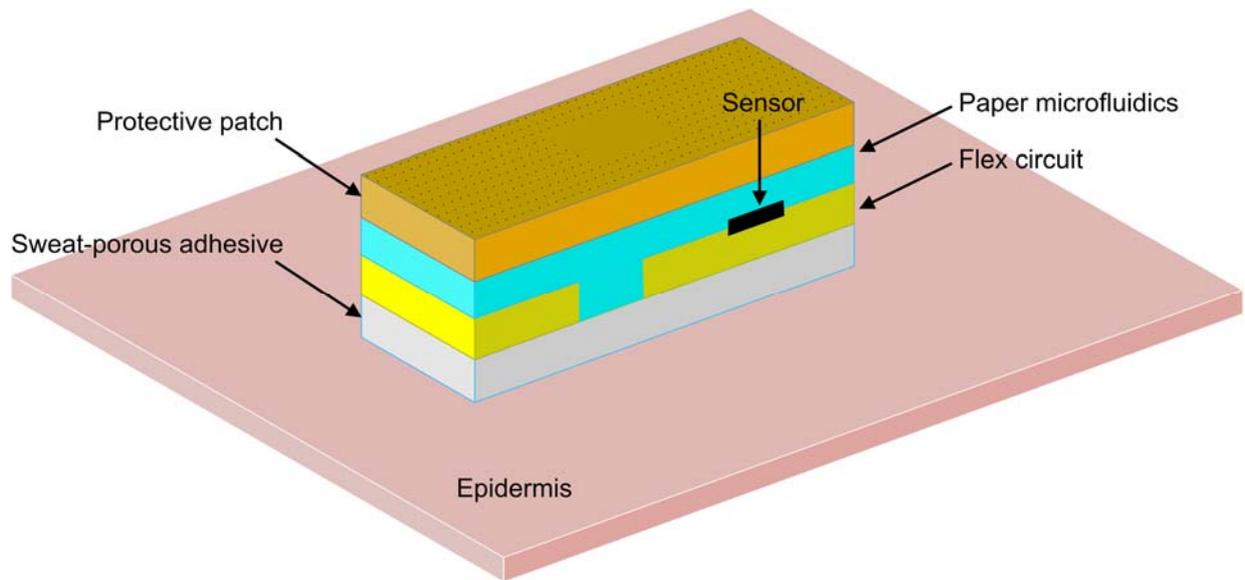

Figure 7: The engineered patch able to collect and analyze the skin sweat in real time. Indeed, the ions contained into the sweat drops are processed by the onboard chip and finally the relevant health information is transmitted to an external device.

An interesting strand of research on bacteria-based communications consists of making use the quorum sensing mechanisms used by bacteria for coordinating colonies. It is a decentralized process, which allows bacteria to estimate their density and to regulate their behavior for obtaining some desired macroscopic effects. An example of a coordinated task is the emission of light signals by the synthesis of the Green Fluorescent Protein (GFP). This mechanism could be used for establishing communications from implanted nanomachines to the outside world [113]. By encoding information through the concentration of Acyl Homoserine-Lactone (AHL) molecules, it is possible to trigger each bacterium to produce GFP with an intensity that depends on the concentration of assimilated AHL molecules. This feature could also be added also in bacteria that normally does not emit GFP, through the configuration of their plasmid [113].

A recent research, lead by the Google X project, proposes the use of wearable devices equipped with a magnet able to attract and count the magnetic nanoparticles released in a body [114]. In this way, nanoparticles can release the collected information, for example about a detected cancer or any other diseases, to the wearable device. The monitoring nanoparticles may be introduced in the human body by taking a pill. Google researchers intends to obtain a proactive medicine, rather than reactive, through a monitoring system of the physiological conditions, integrated in the Google Fit Platform [115].

Finally, Table 1 summarizes the main enabling technologies for the applications mentioned above (associated with rows), showing for each one the available implementation options (associated with columns), ranging from the FRET/QD to the protein-based. This table aims to provide a global picture of the current state of the art technologies that can be used for implementing the specific medical applications identified in our taxonomy. For each application reported in Table 1 we summarize the most promising implementation options illustrated so far. It is worth to note that Table 1 represents the current state of the art. The reported information does not imply that some of the presented technologies in the near future cannot have another medical application.

For what concerns the imaging techniques used for detecting both diseases and other specific markers, the FRET technology and QDs provide interesting solutions, since external devices can easily measure the intensity of light signals. Also antibodies are good implementation options to advanced imaging techniques. For instance, they allows implementing the MRI virtual biopsy technique by using a nanoconjugate carrying MRI contrast agent and antibodies recognizing tumor-specific markers. A similar method is used with chemical antibodies (aptamers). Disease detection can be performed in different ways, including the use of engineered bacteria sensible to specific disease (e.g attracted by tumors lysates), and the use of molecular communication receivers, consisting of nanomachines, which expose surface receptors compliant with cytokines and antibodies released from damaged tissues, infected cells, and immune system cells. Personalized diagnosis can be performed by analyzing the release of proteins that reflect the health condition of a patient. On the basis of release pattern observed, it is possible to build an in-silico model for predicting evolution of diseases. By using nanomachines as actuators (nanorobots) damaged tissues and organs can be repaired from the inside of the human body. This is commonly included in the field of nanosurgery applications. Another instance of medical applications is the targeted release of drugs, which can reduce the amount of medicaments required to treat a disease. Engineered bacteria, viruses, and nanomachines can release drugs in a controlled fashion, thus implementing a bio cyber-physical system. In addition, even modified antibodies and proteins can be released in order to trigger specific behaviors on the target sites. More specifically, it has been demonstrated that the release of specific proteins and growth factors could control the tissue engineering process. Finally, it could be sufficient to trigger the immune system response to remove pathogens and disease conditions more efficiently than it happens spontaneously. This could be obtained by the coordinated action of engineered bacteria, antibody, and viruses, that, after the recognition of an external threat, could release specific molecules (proteins) or RNA strands which alert the immune cells.

| Medicine applications | Implementation options through molecular communications | | | | | |
|---|---|---|---|---|---|---|
| | FRET/QD | Bacteria | Virus | Antibody | Nanorobot | Protein-based |
| Imaging | x | | | x | | |
| Disease detection | | x | | x | | x |
| Personalized diagnosis | | | | | | x |
| Nanosurgery | | | | | x | |
| Drug delivery | | x | x | x | x | x |
| Tissue engineering | | | | | | x |
| Immune system activation | | x | x | x | | x |

Table 1: Implementation options for main medical applications.

## 5. More visionary applications

A novel cancer therapy envisages the use of DNA nanorobots inoculated into a diseased patient. The nanorobots, which are developed in order to not cause an immune response, can perform simple surgery on diseased cells. Moreover, these nanorobots could be combined with toxic drugs for an effective delivery. By means of DNA origami and molecular programming, the nanorobots can discover and kill cancer cells, taking advantage from the reciprocal coordination, generating logical outputs [116].

Molecular communications could also support to the treatment of pain. Indeed, each year many people suffer the pain effects, which leads to work disability by causing a loss in productivity. The pain is a perception, so the idea is to block this perception. The traditional approach makes use of powerful morphine-based painkillers that cause severe side effects on the body, including nausea, vomiting, and constipation. Opioids block pain by binding to 'mu-receptors' in the brain. However, they also bind to mu-receptors in the bowels, and this causes constipation. People with chronic pain need to think outside of the opioid/narcotic pain box. There are also treatments besides narcotics to help chronic pain such as antidepressants. Depending on the type of pain, even some cardiac drugs are effective. For these reasons, an effective contribution on the treatment of pain could rise from the monitoring of the pain interest region with probe based monitoring, optogenetic light switch, and bacteria releasing localized drugs via the optogenetic activation. By such a system, the drug delivery will be determined through a personalized process, driven by optimization techniques and computing. It will be targeted to the specific needs, tolerance to drugs, pain extension, life style (e.g. smoke, alcohol, …). In this way, the treatment will focus on pain type and extension of the interested area. This will improve the quality of life, since no drug flooding through the whole body will be allowed. Psychological issue typically determined by pain treatment will be avoided or minimized. Moreover, the treatment could happen at home, by minimizing medical interaction besides the initial configuration and triggering. This means a significant cost reduction for the national health systems, due to both the reduced amount of drug used and effectiveness of treatment. The medical personnel could be also alerted through smartphone driven by the implanted controller in case the results of treatment is not satisfactory. As presented above, both mobile sensors and actuators (drug releasers) will be bio-compatible devices, degradable through specific enzymes, avoiding the activation of the immune response and any other side effects. This approach could be also a significant component of other treatments where significant pain is expected, thus improving effectiveness and acceptability by patients.

Future research directions in the monitoring of the biological parameters can envisage the analysis of the circulating RNA naturally released by healthy and diseased cells through the exosomes [34]. In more detail, it is known that cancer cells frequently release modified RNA strands that could be collected by bionanomachines (e.g. engineered bacteria, nanorobots, smart probe receptors, and so on), which are previously configured with the 'original' copy of the patient DNA. In this way, the captured RNA strands are compared with the healthy version of the DNA allowing the detection of any not matching pattern. This could be an alternative way for cancer monitoring and identification, without any loss of generality. In fact, also the non tumoral cells could suffer a degradation/overwriting of their nuclear DNA, caused by malicious bacteria and viruses. Indeed, many viruses deploy mechanisms designed to evade pathogen-sensing pathway, in order to spread the viral infection [117], enclosing their RNA into secreted vesicles and exosomes. Such monitoring system could also give information about the nature and identity of such viruses, simplifying not only the diagnosis of the disease but also triggering the release of medicaments and drugs.

A further promising technology, which can promote innovative health care applications, consists of using the so-called clustered, regularly interspaced short palindromic repeats (CISPR). They are DNA sequences containing short repetitions of base sequences, separated by short segments of DNA, which have been obtained from previous exposures to viruses [118]. They can be found in many bacteria and most of archaea, which are single-celled prokaryotic microorganisms. CRISPRs constitute a sort of "genetic memory" used to reject new, returning, and ever-present invading DNA molecules. CRISPRs have been identified as the underlying mechanism of a genetic interference pathway that limits gene transfer and protect cells from bacteriophages and conjugative plasmids. In summary, they implement an efficient antiviral defense mechanism, acting like an immune system. Furthermore, the CRISPR interference can be even reprogrammed in order to reject the invading DNA molecules that have not been still encountered. Potential health-care applications are numerous, and are essentially related to the combined use of molecular communications and CRISPR to rewrite human genes for obtaining innovative treatments. Potential targets of CRISPR-based treatments are genetic diseases, such as cystic fibrosis and sickle-cell anemia, which are caused by single base pair mutations [119].

Finally, another important application field for the health monitoring is given by the communicome, which is the technique that control a large amount of communication factors in the plasma of patients. The cytokines, chemokines and growth factors, which are members of the molecular network, may have different expression levels in patients with neurological diseases compared with those found in healthy people. The measures performed on both healthy people and Alzheimer's disease patients show that it is possible to identify small groups of factors that give a sort of signature of the disease [120]. Using proper nanomachines, this non-invasive technique is exploitable in plasma or other fluids over the course of a neurological disease and may provide information about pathophysiological processes.

In spite of the huge research effort made in recent years, there are still a lot of open challenges on this field. At first, the development of a complete nanomachine is yet tricky. Indeed, the design of such devices involves multidisciplinary expertise in the biological, medicine and engineering fields. In the nature, similar devices are available but forcing them to perform the desired behavior requires a high customization capability of the intracellular compartments and of the cellular physiology. Nanotechnologies have to reach a degree of maturation, which allows, in the next years, the development of bio-hybrid or even full hybrid devices. Indeed, current researches are based on several subunit that are yet to come, such as the elaboration unit of the nanomachines, which has to process the collected information and take decisions based on these. Several proposals on biological logic gates have been done, but their physical implementation is still in its infancy, and currently unavailable on large scale. We could expect that such elaboration units have to be enough powerful to accomplish all the desired tasks, but also tiny and with low energy consumption. In more detail, the last issue deserves attention, since the energy harvesting in a biological environment is a complete new research field, which could take inspiration from the ATP cycle in cells. In this way, it could be possible to obtain chemical energy, which could be exploited both for processing and for motion and other core functions. In case of active motion, this could be obtained by one or more artificial flagella, which rotates at the required speed acting like a propeller, thus supporting both motion and direction control.

Also information storing is a process that consumes energy and space in a nanomachine. Thus, storing capacity and the rate of storing require special attention. One interesting proposal consists of DNA storing, since it is a very reliable molecule, able to store a very high amount of information for a very long time, through base pair coding. How to control this process in a hybrid device is still an open issue.

Communication units have still to be suitably designed at physical level. Indeed, the ligand-receptor binding process, and also the exosome and endosomal release and uptake, have to be induced and controlled to obtain the desired behavior. Moreover, molecular communications require a huge amount of molecules for each transmitted message and, for this reason, the released molecules are to be considered a finite resource that has to be continually replenished, locally produced, and/or recycled. In nature, the biological cells accomplish this task by recycling the internal and absorbed proteins, assembling the molecular chains by means the mRNA sequence and by using complex molecular machines, such as ribosomes. These units are very complex, and a complete artificial ribosome requires a huge effort in the nanotechnology field.

On the communication side, there are many open challenges, such as the rate control strategies, that are able to control the emission of information molecules according to the environmental conditions. Moreover, the communication scenarios analyzed so far are very simple and complete communication protocols able to handle the multi access interference are yet to come. This is a required step to obtain concurrent transmission between multitudes of nodes in the same channel. In addition, also the mechanisms able to increase the reliability on the received information are missing, on both among nanomachines and with external interfaces.

Since the analyzed scenarios are essentially biological, the biocompatibility of such devices have to be analyzed carefully, in order to avoid any side effects on tissues and organs. Moreover, these devices have to be able to degrade themselves naturally when their services are no longer required. A reliable shutdown process has to be developed in order to tear down all the connections and to stop all actuation services [121].

## 6. Conclusion

This paper surveys the medical applications being developed by leveraging on molecular communications. In addition to showing that new significant progresses have been achieved in the fields of diagnosis and treatment, we aim at stimulating further research in this promising and challenging area. In this regard, we have stressed the role of interdisciplinary research and how in-vitro and/or in-silico joint experiments can be used for assessing most of the techniques illustrated, essentially for discovering any unexpected phenomena.

Based on current achievements, we have both proposed a taxonomy of medical applications and identified, for each one, the existing implementation technologies.

Finally, we have explored a number of more advanced and visionary medical applications, along with relevant open research issues relevant to molecular communications, with an attempt of identifying future and more ambitious applications, with an expected significant impact on society.

## Acknowledgement

This work has been supported by EU project H2020 FET Open 665564 - CIRCLE (Coordinating European Research on Molecular Communications).